\documentclass[prb, showpacs,twocolumn]{revtex4}
\usepackage{bm}
\usepackage{graphicx}%
\begin{document}
\title
{Relaxation of superflow in a network: an application to the
dislocation model of supersolidity of helium crystals}
\author{D.\,V.\,Fil$^{1}$ and  S.\,I.\,Shevchenko$^2$}
\affiliation{%
$^1$Institute for Single Crystals, National Academy of Sciences of
Ukraine, Lenin Avenue 60, Kharkov 61001, Ukraine\\ $^2$B.\,Verkin
Institute for Low Temperature Physics and Engineering, National
Academy of Sciences of Ukraine, Lenin Avenue 47 Kharkov 61103,
Ukraine}

\begin{abstract}
We have considered the dislocation network model for the
supersolid state in $^4$He crystals. In difference with uniform
two-dimensional and three-dimensional systems,  the temperature of
superfluid transition $T_c$ in the network is much smaller than
the degeneracy temperature $T_d$. It is shown that  a crossover
into a quasi superfluid state occurs in the temperature interval
between $T_c$ and $T_d$. Below the crossover temperature the time
of decay of the flow increases exponentially under decrease in the
temperature. The crossover has a continuous character and the
crossover temperature  does not depend on the density of
dislocations.
\end{abstract}

\pacs{67.80.bd}

\maketitle
 Forty years ago Andreev and Lifshitz \cite{1} predicted the
possibility of realization of a ground state of quantum crystals
with the number of sites larger than the number of atoms. Empty
sites in such a state are called zero-point vacancies.
In crystals consisting of Bose atoms zero-point vacancies behave
as Bose quasiparticles and at low temperatures they become
superfluid ones. Therefore the phenomenon predicted by Andreev and
Lifshitz is called the supersolids. In 1970s the problem of
supersolids was addressed a number of theoretical papers
\cite{2,3,4,5,6}, but numerous experimental efforts to discover
this phenomenon failed (see Ref. \onlinecite{7}). The situation
has changed completely after the experiment of Kim and Chan
\cite{9} in which a nonclassical rotational inertia (NCRI)  of
$^4$He crystal was
observed below certain critical temperature. 
NCRI
means that the $^4$He
crystal does not rotate as a rigid body, but certain (superfluid)
fraction of atoms is decoupled from the rotation. Further
experiments \cite{10,11,12,13,14} confirmed the NCRI effect. 
It was established \cite{10,11,12,13,14,bb} that the NCRI is not
an intrinsic property of $^4$He crystals. The amount of fraction
decoupled from the oscillations (rotation) depends considerably on
the degree of disorder in the crystal lattice, in particular, it
becomes much smaller after annealing. A correlation between
supersolid features and disorder was also observed in the direct
flow experiment \cite{rh}.
A common explanation of the correlation between disorder and
supersolid behavior is that zero-point vacancies may emerge only
at extended crystal lattice defects, presumable, in dislocation
cores. The idea on superfluid behavior of dislocations was put
forward in Ref. \onlinecite{sh1} (originally, in a context of
description of anomalous plastic properties of parahydrogen
\cite{e1}). In Ref. \onlinecite{15}  this idea was applied for the
explanation of NCRI. In \onlinecite{15} and in the following
 \cite{16} papers superfluid properties of dislocations were
demonstrated by the first principles Monte Carlo calculations. A
combined (dislocation network plus bulk) mechanism of
supersolidity was considered in Ref. \onlinecite{toner}. The role
of dislocations in this mechanism 
is similar to one
for the enhancement of superconductivity by dislocations
\cite{nab}.


In this Rapid Communication we study the specifics of the
superfluid transition in the network of one-dimensional (1D) wires
and arrive at the following conclusion.
 The temperature of the superfluid transition in the network is
quite small and it depends on the length of segments of the
network. At the same time,  a crossover into a quasi superfluid
state (a state with exponentially large time of relaxation of a
flow) takes place at much larger temperatures, and the crossover
temperature does not depend on the length of the segments. It is
just the behavior observed in torsion experiments where the
transition is continuous, and it occurs in the temperature
interval independent of the NCRI fraction.

The problem of the superfluid transition in a dislocation network
was studied in Ref. \onlinecite{sh2}. It was shown that in the
two-dimensional (2D) network the critical temperature is
proportional to the inverse length of the segment. Let us derive
the result of Ref. \onlinecite{sh2} using the following simple
arguments. For a 2D uniform medium the
Berezinskii-Kosterlitz-Thouless (BKT) critical temperature is
determined by the
 equation   $T_{c}=\pi\hbar^2 n_{s2}(T_c)/2m$, where
$n_{s2}$ is the two-dimensional superfluid density. The BKT
transition is connected with the energy of a vortex pair diverges
logarithmically at large distances between the vortices. At such
distances the difference between a uniform medium and a network is
not important. Therefore, the critical temperature for the network
is  given by the BKT equation in which the uniform density
$n_{s2}$ is replaced with the average 2D superfluid density
$\bar{n}_{2}$. The latter quantity can be expressed through the 1D
density of zero-point vacancies $n$ and the length of the segment
of the network $l$. For the quadratic network $\bar{n}_{2}=2n/l$,
and the critical temperature is $T_c=\pi\hbar^2 n/ml$  ($m$ is the
effective mass of zero-point vacancies).

In  Ref. \onlinecite{18} a microscopic model of Bose-Einstein
condensation (BEC) of non-interacting zero-point vacancies in a
three-dimensional (3D) network was considered. It was shown that
the BEC temperature depends significantly on the transparency of
vertices (intersections of dislocations), and the highest value,
reached at high transparency, is $T_{\rm BEC}\approx 3\hbar^2
n/2ml$, so the superfluid critical temperatures in 3D and 2D
networks are of the
same order. 

The critical temperature $T_c$ in a network is much smaller than
the degeneracy temperature for the 1D Bose gas of vacancies in a
segment $T_d=\hbar^2 n^2/m$ (the small parameter is $(nl)^{-1}\ll
1$). In this respect the network differs from uniform 2D and 3D
systems, where $T_c\approx T_d$. One can expect that the network
may demonstrate quasi superfluid behavior in a wide range of
temperatures between $T_c$ and $T_d$.

To be more specific we consider a 2D network. In a uniform 2D
system  the vortex-antivortex pairs  unbind above the BKT
transition and vortices of opposite vorticities can move
independently from each other. If a given vortex crosses the
system in a direction perpendicular the flow, the superfluid phase
difference along the flow changes on $2\pi$.
In a uniform system a motion of a
vortex across the flow is caused by the Magnus force and the
viscous friction between the vortex and the normal component. The
network is a multiple connected system, the vortices correspond to
circular currents, and they are pinned to  given plaquettes. The
vortex centers cannot move freely, but they can jump from one
plaquette to another. This process becomes possible due to an
emergence of phase slip (PS) centers at the segments.

The theory of dissipation of supercurrent in 1D channels based on
the idea of emergence of PS centers was put forward by Langer and
Ambegaokar \cite{ab}. The main shortage of the theory \cite{ab} is
that it does not yield the pre-exponential factor in the
expression for the relaxation time. To obtain this factor one
should consider the dynamics of transition of the system over the
potential barrier under appearance of a PS center. This problem
was solved in Ref. \onlinecite{mh} on the base of the
\textit{diffusive} (time-dependent Ginzburg-Landau) equation. We
solve the problem with the use of the \textit{wave}
(Gross-Pitaevskii) equation that describes an essentially
different physical mechanism of relaxation of the supercurrent.

If the vortex is already present in a given plaquette the PS of
the proper sign provides annihilation of that vortex and creation
of a vortex of the same vorticity in the neighbor plaquette. A
vortex may jump to any neighbor plaquette, but there is a
preferable direction of such jumps in a system with a flow. Let
us, for simplicity, consider a regular quadratic network with a
flow directed parallel to the segments (say $x$ direction). If the
vortex is centered in a given plaquette the superfluid velocities
in the segments that form this plaquette read as $v_A=v_v+v_s$,
$v_C=v_v-v_s$, $v_B=v_D=v_v$, where A and C stand for the segments
oriented along the $x$ axis, while B and D - for the segments
oriented along the $y$ axis (perpendicular to the flow). Here
$v_s$ is the flow velocity, and $v_v=\pi\hbar/2m l$ is the
contribution caused by the vortex. The frequency of PS in a
segment is proportional to its length and it is a function of the
superfluid velocity $\nu_i=l f(v_i)$. One can see that
$\nu_B=\nu_D$ and in average the vortices do not move in the $x$
direction (the direction of the flow). The PS frequencies for two
other segments differ from each other and one extra jump in the
perpendicular to the flow direction
gains with the
frequency
\begin{equation}\label{3}
    \Delta\nu= \nu_A-\nu_C= \alpha l v_s,
\end{equation}
where $\alpha=2 f'(v)|_{v=0}$ (here we imply the limit of small
superfluid velocities). In the network of a rectangle shape of
area $S=L_x\times L_y$ a vortex crosses the system with the
frequency $\nu_{cross}= N_v \Delta\nu l/L_y$, where $N_v=n_v S$ is
the total number of unbound vortices ($n_v$ is the vortex
density). Each cross  lowers the phase gradient on
 $2\pi/L_x$ that changes  the flow
 velocity on $\Delta v_s= -2\pi\hbar/m L_x$.
The equation for $v_s$, written in the differential form, reads as
\begin{equation}\label{4}
    \frac{d v_s}{d t}=\nu_{cross}\Delta v_s=  -\frac{2 \pi\hbar\alpha  n_v l^2}{m}
    v_s.
\end{equation}
 The solution of Eq. (\ref{4}) is $v_s=v_{s0}e^{-t/\tau}$, where
$\tau=m/2 \pi\hbar\alpha n_v l^2$ is the decay time. To compute
$\alpha$ one should specify the mechanism of PS. We will describe
the gas of zero-point vacancies in the dislocation core  as
 a weakly non-ideal 1D  Bose gas with a complex
condensate wave function (order parameter) $\Psi(x,t)$ that
satisfies the Gross-Pitaevskii (GP) equation. The 1D GP equation
with a repulsive point interaction has an exact solution that
corresponds to a dark soliton. The dark soliton is a rarefaction
that moves with a constant velocity $u$.

The dark soliton is described by the  function \cite{a1}
\begin{equation}\label{5}
\Psi(x,t)=\sqrt{\tilde{n}}
    \left[\sqrt{1-\frac{u^2}{c^2}}
    \tanh\left(\sqrt{1-\frac{u^2}{c^2}}\frac{x-u
    t}{\xi}\right)+i\frac{u}{c}\right].
\end{equation}
Here $\xi=\hbar/mc$ is the coherence length, $c=\sqrt{\gamma
{n}/m}$ is the sound
 velocity, $\gamma$ is the interaction constant,
 $\tilde{n}=n(1-2(\xi/l) \sqrt{1-u^2/c^2})^{-1}$
 is the renormalized density
   (renormalization is the consequence of the conservation of the
  total number of zero-point vacancies).
 The energy of the soliton reads as
\begin{eqnarray}\label{9e}
    E_{0}=\int_0^{l} d x \left[\frac{\hbar^2}{2
    m}\left(\frac{d \Psi}{d
    x}\right)^2+\frac{\gamma}{2}(|\Psi|^4-n^2)\right]=\cr
    \frac{4}{3}\hbar n c
    \left(1-\frac{u^2}{c^2}\right)^{3/2}.
\end{eqnarray}
The soliton momentum $p$ can be found by integration of equation
$dp=dE_0/u$:
\begin{equation}\label{8}
    p=-2\hbar n \left(\frac{u}{c}\sqrt{1-\frac{u^2}{c^2}}+
    \arcsin\frac{u}{c}\right)+C,
\end{equation}
where $C$ is the constant of integration. To determine $C$ one can
take into account that the soliton momentum is the difference of
the momentum of the system with  and without the soliton. The
soliton emerges with the velocity $u =+c-0$ or $u=-c+0$ and  at
such $u$ its momentum should be equal to zero. One can see that
two conditions $p_{u=\pm c}=0$ yield two different integration
constants $C_{\pm}=\pm\hbar n \pi$. Therefore, one should consider
two species of the solitons (the "+" and "-" ones) with the
momenta $p_{\pm}$ defined by Eq. (\ref{8}) with $C=C_{\pm}$.

It is important to emphasize that these two species correspond two
physically distinct solitons. According to Eq. (\ref{5})  the
phase at the soliton has the additional shift $\Delta \varphi(u) =
-2 \cot^{-1}({u}/{\sqrt{c^2-u^2}})$. In any multiple connected
system the phase satisfies the Onsager-Feynman quantization
condition, and the appearance of a soliton should be accompanied
by a change in the net velocity: $v=v_0+\Delta v_\pm(u)$. The
function $\Delta \varphi(u)$ is discontinuous at $u=0$ with the
jump equal $2\pi$. The functions $\Delta v_\pm(u)$ should be
continuous because small variation in $u$ cannot result in a
finite change in the net velocity. Since $\Delta v_+(+c)=0$ and
$\Delta v_-(-c)=0$, the function $\Delta v_+(u)\ne\Delta v_-(u)$.
For instance, for a 1D ring with the perimeter $l$  we find
$\Delta v_{\pm}=\hbar(\pm\pi-2\arcsin({u}/{c}))/{m l}$ (we take
into account that $\Delta\varphi+ml\Delta v_\pm/\hbar=0 \ {\rm
mod}\ 2\pi$). Thus, solitons of distinct species differ from each
other by the change in the net velocity they induce.

 The soliton may
change its velocity due to the interaction with phonons and
impurities. Since the soliton changes the net velocity in a
continuous way, it is more consistent to consider PS as an entire
process of creation of the soliton at $u=+c$ and its annihilation
at $u=-c$, or vise versa. The first possibility corresponds to the
"+" solitons, and the second one - to the "-" ones.

To obtain the frequency of PS we consider solitons as classical
particles whose distribution functions $f_{\pm}(p,t)$ satisfy the
Fokker-Planck equation
\begin{equation}\label{12}
    \frac{\partial f_{\pm}}{d t}=-\frac{\partial s_{\pm}}{\partial
    p},
\end{equation}
where $s_{\pm}$ are the  soliton fluxes in the momentum space.
They read as
\begin{equation}\label{13}
    s_{\pm}=A_{\pm}f_{\pm}-B_{\pm}\frac{\partial}{\partial p}
    f_{\pm}.
\end{equation}
The coefficients $A_{\pm}$ and $B_{\pm}$  satisfy the  relation
$A_{\pm} f_{0,\pm}-B_{\pm}
\partial f_{0,\pm} /\partial p=0$, where
$f_{0,\pm}=\exp(-E_{\pm}/T)$ is the equilibrium distribution
function, and $E_{\pm}=E_0+p_{\pm}v$ are the soliton energies at
nonzero net velocity. Here we consider the case of slow
relaxation. In this case one can neglect the explicit  time
dependence of $f_{\pm}$   and consider the fluxes $s_{\pm}$  as
constant quantities. Using the relation between the coefficient
$A$ and $B$ we rewrite Eq. (\ref{13}) in the form
\begin{equation}\label{14}
    s_{\pm}=-B_{\pm}f_{0,\pm}\frac{\partial}{\partial p}\left(
    \frac{f_{\pm}}{f_{0,\pm}}\right).
\end{equation}
 The distribution of the solitons with small $p$  is
close to equilibrium one. The solitons with $p_{\pm}\to \pm 2\pi
\hbar n$ emerge only due to nonzero $s_{\pm}$. To attain such a
momentum the soliton should overcome the energy barrier $\Delta
E=4\hbar n c/3$. Therefore the fluxes are small and the
distribution functions $f_{\pm}$ at $p=\pm 2\pi \hbar n$ are much
less than the equilibrium ones. The integration of (\ref{14}) with
the boundary conditions $(f/f_0)|_{p=0}=1$ and $(f/f_0)|_{p=\pm
2\pi \hbar n}=0$ yields
\begin{equation}\label{15}
    s_{\pm}=\left(\int_0^{\pm
2\pi \hbar n}\frac{d p}{B_{\pm}f_{0,\pm}}\right)^{-1}.
\end{equation}
Here we imply the case of small temperatures $T\ll \Delta E$ and
small net velocities $v\ll c$. Then in the leading order the
integral (\ref{15}) is evaluated as
\begin{equation}\label{16}
    s_{\pm}=\pm
\frac{B_{0,\pm}}{4\hbar n} \left(\frac{2 \hbar n c}{\pi
T}\right)^{1/2} \exp\left(-\frac{4\hbar n c}{3 T}\mp
\frac{\pi\hbar n
    v_{\pm}}{T}\right),
\end{equation}
where $B_{0,\pm}$ is the 'diffusion" coefficient $B_{\pm}$  at
$u=0$.  Using the exact form of $f_0$ one finds that
$B_{\pm}=-A_{\pm}T/(u+v)$. The coefficient $A=d p/dt$ is just the
viscous friction force acting on solitons. At small velocities
this force is proportional to the velocity of the soliton motion
relative the normal component $A=-\eta (u+v)$, that yields $B=\eta
T$, where $\eta$ is the friction coefficient (under assumption
that $\eta$ is the same for "+" and "-" solitons, the coefficients
$A_\pm=A$ and $B_\pm=B$ are the same as well).

The soliton distribution functions  are normalized by the
condition $n_{sol}=(1/2\pi\hbar)\int f dp$, where $n_{sol}$ is the
soliton density. The quantities $s_\pm$ are
 the fluxes in the direction of larger momenta $p$
(negative sign of $s_-$ means that actual direction of the flux is
the opposite one). The difference of their modules determines the
frequency of PS's:
 \begin{equation}\label{17}
    \nu=\frac{(|s_-|-s_+)l}{2\pi \hbar}.
\end{equation}
Using Eqs. (\ref{16}) and (\ref{17}) one finds the parameter
$\alpha$
 and obtains the following expression
for the decay time
\begin{equation}\label{19}
    \tau=\tau_0 \frac{1}{n_v l^2}  \left(\frac{
    T}{2\pi\hbar n c}\right)^{1/2}\exp\left(\frac{4\hbar n c}{3
    T}\right),
\end{equation}
where $\tau_0=m/\eta$. According to (\ref{19}) the crossover
temperature is $T_0= 4\hbar n c/3$. For weakly non-ideal Bose gas
$T_0/T_d\sim \sqrt{\gamma n/T_d}\ll 1$ and $T_0/T_c\sim l/\xi \gg
1$, so $T_c\ll T_0\ll T_d$.

 To evaluate the  parameter $\tau_0$ we consider the
friction connected with the interaction of the solitons with
phonons. The dark soliton is the exact solution of the 1D GP
equation and it does not interact with phonons in 1D. But the
dislocation core is not a strict 1D system. It is a quasi-1D
system with a small, but finite cross-section.  Such a system is
described by an effective 1D GP equation with an additional higher
order in $\Psi$ interaction term \cite{sh}: $i\hbar
\partial\Psi/\partial t=-(\hbar^2/2m)\partial^2\Psi/\partial
x^2+\gamma|\Psi|^2\Psi -\gamma_1|\Psi|^4 \Psi$. Due to such a term
the reflection coefficient $R\sim (\gamma_1 n/\gamma)^2$ for the
phonons that scatter on soliton is nonzero \cite{sh}.


The parameter $\gamma_1 n$ is evaluated as $\gamma_1 n\sim \gamma
(r_\perp/\xi)^2$, where $r_\perp$ is the radius the supefluid
channel. As was shown in Ref. \onlinecite{sh}, at small $u$ (and
$v=0$) the time derivative of the soliton momentum is given by the
expression $\dot{p}=-\eta u= - C (m R T/\hbar)u$, where the
numerical factor $C\sim 1$. It yields $\tau_0\sim (\hbar/T)
(\xi/r_\perp)^4$. The scale of $\tau_0$ is determined by the
quantity $\hbar/T$ that is of order of $10^{-10}$ c$^{-1}$ for $T=
0.1$ K.

It is necessary to note one important point.  One could think that
since the GP equation we use is invariant with respect to
translations it is impossible to describe the relaxation of
superflow in the GP approach. But our approach is basically the
same as commonly used for the computation of forces that act on a
vortex in a 3D superfluid caused by its interaction with phonons
and rotons (see, for instance Ref. \onlinecite{sonin}). In the
latter case the GP approach is applied for the computation of
amplitudes of scattering  of phonons on vortices and for finding
the momentum flux  over the cylindrical surface around the vortex
line. In such a way one can obtain the rate of transfer of the
momentum from a vortex to phonons, i.e.., to the normal component,
which is assumed to be in equilibrium with the environment (walls,
substrate, etc.)

The specific of the 1D system is that for the GP equation with
only cubic interaction term the phonons do not interact with
solitons and there is no transfer of the momentum from solitons to
the normal component. The fifth-order term switches on that
interaction, and the momentum transfers from solitons to 1D
phonons of the network. 1D phonons interact with bulk phonons and
due to such an interaction the normal component
in the network remains in equilibrium with the crystal (since the
relaxation time in the phonon subsystem is much smaller than
$\tau$ given by Eq. (\ref{19})).
%
Eventually, the momentum obtained from solitons is transferred to
the crystal.
Here we do not describe explicitly the mechanism of such a
transfer, but  just imply that the normal component is in
equilibrium with the environment.

Estimating (\ref{19}) we obtain that  the decay time is of order
of few seconds at $T \approx 0.04 T_0$ and it is of order of an
hour at $T \approx 0.03 T_0$. We note that at very small
temperatures quantum jumps between states with different vortex
configurations may become important (for 1D rings the quantum
jumps were studied in Refs. \onlinecite{23,24}). We estimate the
quantum correction to the $\tau$ is $\tau_0\propto e^{n\xi}$ and
this correction becomes important at $T\lesssim \gamma
n=T_0/n\xi\ll T_0$. The quantum correction results only in a
modification of the law of increase in the relaxation time under
lowering of temperature and its accounting  should not change the
main conclusion on the emergence of the quasi superfluid state in
the network well above the critical temperature.

In this Rapid Communication we have considered quasi superfluidity
in a 2D network. The situation in a 3D network should be
qualitatively the same. In the latter case superflow may decay due
to expansion or shrinking of vortex rings. The mechanism of
expansion(shrinking) of vortex rings in a 3D network is basically
the same as vortex motion in a 2D network: both of them are
connected with the phase slips in segments of the network.

In conclusion, we note that the results obtained can be also
applied to multiple connected Bose-Einstein condensates of
rarefied alkali gases in optical lattices \cite{ol}, where the
quasi superfluid state can be observed directly.

We are grateful N.V.Prokof'ev, B.V.Svistunov and G.V.Shlyapnikov
for the discussion. This Work was supported in part by the CRDF
Grant No. UKR2-2853. We also acknowledge LPTMS, University
Paris-Sud, where part of this work was done, for the hospitality.

\end{document}